\global\def\draftcontrol{0}

   \def\versionno{ alpha prime fluctuations complete -- draft   }
\catcode`\@=11

\expandafter\ifx\csname draftcontrol\endcsname\relax\global\def\draftcontrol{0}
\fi

{\count255=\time\divide\count255 by 60
\xdef\hourmin{\number\count255}
\multiply\count255 by-60\advance\count255 by\time
\xdef\hourmin{\hourmin:\ifnum\count255<10 0\fi\the\count255}}
\def\draftdate{\number\month/\number\day/\number\year\ \ \ \hourmin }

\newcommand\makepapertitle{\par
  \begingroup
    \renewcommand\thefootnote{\@fnsymbol\c@footnote}%
    \def\@makefnmark{\rlap{\@textsuperscript{\normalfont\@thefnmark}}}%
    \long\def\@makefntext##1{\parindent 1em\noindent
            \hb@xt@1.8em{%
                \hss\@textsuperscript{\normalfont\@thefnmark}}##1}%
     \newpage
     \global\@topnum\z@   % Prevents figures from going at top of page.
     \@makepapertitle
     \thispagestyle{empty}\@thanks
  \endgroup
  \setcounter{footnote}{0}%
  \global\let\thanks\relax
  \global\let\makepapertitle\relax
  \global\let\@makepapertitle\relax
  \global\let\@thanks\@empty
  \global\let\@author\@empty
  \global\let\@date\@empty
  \global\let\@title\@empty
  \global\let\title\relax
  \global\let\author\relax
  \global\let\date\relax
  \global\let\and\relax
  \def\version{\let\version\@version\@gobble}
}
\def\@makepapertitle{%
  \newpage
   \ifnum\draftcontrol=1 {}
   \version\versionno
   \vskip 3em%
   \else
   \hfill\hbox to 3cm {\parbox{4cm}{\@pubnum}\hss}%
   \vskip 3em%
   \fi
   \begin{center}%
   \let \footnote \thanks
     {\LARGE {\@title}}%
     \vskip 1.5em%
     {\normalsize%\large
       \lineskip .5em%
       \begin{tabular}[t]{c}%
         \@author
       \end{tabular}\par}%
     \vskip 1.5em%
     {\@bstract}%
     \end{center}%
     \vskip 1.5em
     \@date%
   \par
}

\gdef\@pubnum{}
\def\pubnum#1{%
  \gdef\@pubnum{#1}}

\gdef\@bstract{}
\def\Abstract#1{%
  \gdef\@bstract{%
   \parbox{\textwidth-0pc}{%
   \centerline{\bf Abstract}\penalty1000%
\kern.2cm%
\noindent%\abstractfont \baselineskip=12pt
\renewcommand\baselinestretch{1.0}%
{#1}}}
}

\def\ps@paper{\let\@mkboth\@gobbletwo%
     \ifnum\draftcontrol=1
	\def\@oddfoot{\hbox to \textwidth{\tiny \versionno \hfil\tiny\draftdate}%
	\hskip -\textwidth \hbox to \textwidth{\hfil\rm\thepage\hfil}}%
     \else\def\@oddfoot{\hbox to \textwidth{\hfil\rm\thepage\hfil}}
     \fi
     \let\@evenfoot\@oddfoot
}
\def\body{\clearpage
          \pagestyle{paper}
	}
\def\@version#1{\ifnum\draftcontrol=1
\typeout{}\typeout{#1}\typeout{}
\vskip3mm\centerline{\hbox{\fbox{\normalsize{\tt DRAFT -- #1 -- }
                   {\draftdate}}}}\vskip3mm
\fi}
\let\version\@version
\long\def\eqlabel#1{\ifnum\draftcontrol=1
                    \tag@false  % there are some problems with multline without this
                    \tag*{(\theequation) \hbox to -0.2cm{\hspace{0cm}\small{#1}\hss}}
                    \refstepcounter{equation}
                    \edef\@currentlabel{\theequation}
                    \ltx@label{#1}          % use old LaTeX \label instead of new definition
                    \else
                    \label{#1}
                    \fi
                    }
\let\st@bibitem\@bibitem
\let\st@lbibitem\@lbibitem
\ifnum\draftcontrol=1
  \def\@bibitem#1{%
    \st@bibitem{#1}\a@@label{#1}\ignorespaces}
  \def\@lbibitem[#1]#2{%
    \st@lbibitem[#1]{#2}\a@@label{#2}\ignorespaces}
  \def\a@@label#1{%
    \gdef\a@lab{\smash{\normalfont\small#1}}
    \ifvmode
      \if@inlabel
        \global\setbox\@labels\hbox{%
          \llap{\a@lab\let\a@lab\relax
                \kern\@totalleftmargin\kern\marginparsep}%
          \box\@labels}%
      \fi
    \fi}
\fi
\documentclass[12pt,letterpaper]{article}
\usepackage{amsmath,amssymb,array,calc,rotating,epsfig,psfrag}
\usepackage[nosort]{cite}
\ifnum\draftcontrol=1
\fi
\renewcommand\baselinestretch{1.25}
\renewcommand\section{\@startsection {section}{1}{\z@}%
                                   {-3.5ex \@plus -1ex \@minus -.2ex}%
                                   {2.3ex \@plus.2ex}%
                                   {\normalfont\large\bfseries}}
\renewcommand\subsection{\@startsection{subsection}{2}{\z@}%
                                   {-3.25ex\@plus -1ex \@minus -.2ex}%
                                   {1.5ex \@plus .2ex}%
                                   {\normalfont\normalsize\bfseries}}
\renewcommand\subsubsection{\@startsection{subsubsection}{3}{\z@}%
                                   {-3.25ex\@plus -1ex \@minus -.2ex}%
                                   {1.5ex \@plus .2ex}%
                                   {\normalfont\normalsize\it}}
\renewcommand\paragraph{\@startsection{paragraph}{4}{\z@}%
                                   {-3.25ex\@plus -1ex \@minus -.2ex}%
                                   {1.5ex \@plus .2ex}%
                                   {\normalfont\normalsize\bf}}

\numberwithin{equation}{section}

\def\ie{{\it i.e.}}

\def\revise#1       {\raisebox{-0em}{\rule{3pt}{1em}}%
                     \marginpar{\raisebox{.5em}{\vrule width3pt\
                     \vrule width0pt height 0pt depth0.5em
                     \hbox to 0cm{\hspace{0cm}{%
                     \parbox[t]{4em}{\raggedright\footnotesize{#1}}}\hss}}}}

\newcommand\nxt[1]  {\\\fnxt#1}

\def\calc         {{\cal C}}
\def\cald         {{\cal D}}

\def\calf         {{\cal F}}

\def\caln         {{\cal N}}
\def\calo         {{\cal O}}

\def\del          {\partial}

\def\sqr#1#2{{\vcenter{\vbox{\hrule height.#2pt
 \hbox{\vrule width.#2pt height#1pt \kern#1pt
 \vrule width.#2pt}\hrule height.#2pt}}}}

\newcommand{\ft}[2]{{\textstyle{\frac{#1}{#2}}}}

\def\a{\alpha}
\def\w{\omega}
\def\g{\gamma}
\def\hh{\hat{h}}
\newcommand{\qq}{\mathfrak{q}}
\newcommand{\ww}{\mathfrak{w}}
\def\hs{\hat{s}}
\def\hc{\hat{c}}

\catcode`\@=12

\begin{document}

%%%
%%%%%% text starts here
%%%%%%%%%

\title{Transport properties of $\caln=4$ supersymmetric Yang-Mills theory 
at finite coupling}

\pubnum{%
UWO-TH-05/14
}
\date{October 2005}

\author{
Paolo Benincasa$ ^1$ and   Alex Buchel$ ^{1,2}$\\[0.4cm]
\it $ ^1$Department of Applied Mathematics\\
\it University of Western Ontario\\
\it London, Ontario N6A 5B7, Canada\\[0.2cm]
\it $ ^2$Perimeter Institute for Theoretical Physics\\
\it Waterloo, Ontario N2J 2W9, Canada\\
}

\Abstract{
Gauge theory-string theory duality describes strongly coupled
$\caln=4$ supersymmetric $SU(n_c)$ Yang-Mills theory at finite
temperature in terms of near extremal black 3-brane geometry in type
IIB string theory.  We use this correspondence to compute the leading
correction in inverse 't Hooft coupling to the shear diffusion
constant, bulk viscosity and the speed of sound in the large-$n_c$
$\caln=4$ supersymmetric Yang-Mills theory plasma.  The transport
coefficients are extracted from the dispersion relation for the shear
and the sound wave lowest quasinormal modes in the leading order
$\a'$-corrected black D3 brane geometry. We find the shear viscosity
extracted from the shear diffusion constant to agree with result of
[hep-th/0406264]; also, the leading correction to bulk viscosity and
the speed of sound vanishes. Our computation provides a highly
nontrivial consistency check on the hydrodynamic description of the
$\a'$-corrected nonextremal black branes in string theory.
}

%\enlargethispage{1.5cm}

\makepapertitle

\body

\version\versionno

\section{Introduction}
The correspondence between gauge theories and  string theory of Maldacena \cite{m9711,m2}
has become a valuable tool in analyzing near-equilibrium dynamics of strongly coupled 
gauge theory plasma \cite{ss,hs,ne1,ne2,ne3,ne4,ne5,kss,bl1,kss1,bh1,bls,bh2,set,bbs,aby,bkt}. 
The best studied example of strongly coupled thermal gauge theory plasma 
is that of the $\caln=4$ $SU(n_c)$ supersymmetric Yang-Mills theory (SYM). In the large-$n_c$ limit, 
and at large 't Hooft coupling $g_{YM}^2 n_c\gg 1$, the holographic dual description of the 
$\caln=4$  plasma is in terms of near-extremal black 3-brane geometry in type IIB supergravity 
\cite{gkp}. In this case one finds \cite{ne1,ne2,ne4} the speed of sound $c_s$, the shear viscosity 
$\eta$, and the bulk viscosity $\zeta$  correspondingly
\begin{equation}
c_s=\frac{1}{\sqrt{3}}\,,\qquad \eta=\frac{\pi}{8}n_c^2 T^3\,,\qquad \zeta=0\,.
\eqlabel{n4res}
\end{equation}

In a hydrodynamic approximation to near-equilibrium dynamics of hot gauge theory 
plasma there are several distinct ways to extract the transport coefficients \eqref{n4res}.
First \cite{ne1}, the shear viscosity can be computed from the two-point 
correlation function of the stress-energy\footnote{Computation of the 
thermal correlation functions in the dual supergravity description was explained in \cite{ss,hs}.} 
tensor at zero spatial momentum via the Kubo formula 
\begin{equation}
\eta=\lim_{\w\to 0} \frac{1}{2\w}\int dtd\bar{x}\ e^{i\w t}\langle\left[T_{xy}(x),\ T_{xy}(0)\right]\rangle\,.
\eqlabel{kubo}
\end{equation} 
Second \cite{ne2}, the 
diffusive channel two-point retarded correlation function of the stress energy tensor, for example,
\begin{equation}
\begin{split}
G_{tx,tx}(\w,q)=-i \int dt d\bar{x} e^{i\w t-i q z}\theta(t)   \langle\left[T_{tx}(x),\ T_{tx}(0)\right]\rangle
\propto  \frac{1}{i\w-\cald q^2}
\end{split}
\eqlabel{shearchan}
\end{equation} 
has a pole at 
\begin{equation}
\w=-i \cald q^2\,,
\eqlabel{diffpole}
\end{equation}
where the shear diffusion constant $\cald$ is 
\begin{equation}
\cald=\frac{\eta}{s T}\,,
\eqlabel{ddef}
\end{equation}
with $s$ being the entropy density of the gauge theory plasma.
From the thermal field theory perspective it is clear that computation of the shear viscosity via Kubo relation 
\eqref{kubo}, or from the pole of the stress-energy correlation function \eqref{shearchan} (additionally using 
the equation of state to relate \eqref{ddef}) {\it must} give the same result. It is much less obvious that 
such an agreement should persist automatically also on the supergravity side. Thus, we regard  above consistency
of the hydrodynamic description of the black 3-branes in type IIB supergravity as a highly nontrivial check of the 
Maldacena correspondence \cite{m9711} applied to near-equilibrium thermal gauge theories. 

The situation with the sound wave propagation in the hydrodynamic 
limit is similar  \cite{ne4} (though perhaps less dramatic compare with
shear viscosity given the conformal invariance of the $\caln=4$ SYM).      
The speed of sound can be computed from the equation of state as 
\begin{equation}
c_s^2=\frac{\del P}{\del \epsilon}\,,
\eqlabel{cs}
\end{equation}
where $P$ and $\epsilon$ are correspondingly the pressure and the energy density of the 
strongly coupled gauge theory plasma which can be extracted from the thermodynamic 
properties of the black 3-branes \cite{gkp}. Alternatively, all the transport coefficients 
\eqref{n4res} can be read off from the dispersion relation for the pole in the sound wave 
channel two-point retarded correlation function of the stress energy tensor, for example,
\begin{equation}
\begin{split}
G_{tt,tt}(\w,q)=-i \int dt d\bar{x} e^{i\w t-i q z}\theta(t)   \langle\left[T_{tt}(x),\ T_{tt}(0)\right]\rangle\,,
\end{split}
\eqlabel{soundchan}
\end{equation}
as 
\begin{equation}
\w(q)=c_s q -i\ \frac{2q^2}{3 T }\ \frac{\eta}{s}\ \left(1+\frac{3\zeta}{4\eta}\right)\,. 
\eqlabel{sounddisp}
\end{equation}
Again, all these computations point to a consistent picture of a hydrodynamic description of the 
supergravity black 3-branes\footnote{Consistency of hydrodynamic description of more complicated 
examples of gauge theory-supergravity correspondence follows from \cite{bh1,bbs,aby,bgm,bkt}.}. 
 
In this paper we prove that consistent hydrodynamic description of black 3-branes persists even 
once one include leading $\a'$ correction to type IIB supergravity from string theory \cite{cor1,cor2,cor3,cor4}, 
which translates into finite 't Hooft coupling correction on the $\caln=4$ SYM side of the Maldacena duality.    
To appreciate the nontrivial fact of the agreement we point to some features of $\a'$-corrected description of the 
black 3-branes: 
\nxt including leading order $\a'$ correction, the entropy density of the black 3-branes 
differs from the Bekenstein-Hawking formula which relates the latter to the area of the    
horizon \cite{gkt};
\nxt the Hawking temperature of the black 3-branes as well as their equilibrium
thermodynamic quantities, \ie, the entropy, energy and the free energy, receives $\a'$ corrections \cite{gkt,pt};
\nxt unlike the supergravity approximation \cite{gkp}, 
the radius of the $S^5$ of the $\a'$ corrected black 3-brane geometry is not constant \cite{pt}.  

We find that only properly accounting for {\it all} of the above facts one finds a consistent picture of the $\a'$
corrected black 3-brane hydrodynamics. Lastly, we strongly suspect that consistency of the hydrodynamics 
is sensitive to the structure of the $\a'$ corrections in type IIB string theory. Thus our computations can be helpful 
is determining exact structure of such corrections\footnote{We hope to report on this elsewhere.}.

The paper is organized as follows. In the next section we discuss our computational approach and 
present the results. In section 3 we  apply the general computational scheme to the evaluation of the 
dispertion relation of the shear quasinormal mode in black 3-brane geometry without 
$\a'$ corrections. This was previously discussed in \cite{set}, though our approach highlights the use 
of the effective action rather than equations of motion\footnote{Using equations of motion 
is technically much more complicated in the presence of $\a'$ corrections.}.      
In section 4 we discuss the main computational steps leading to the shear and  the sound wave 
lowest quasinormal modes  dispertion relations.

\section{General computational approach and the results}
In the context of  gauge theory-string theory correspondence \cite{m9711}
poles of the finite temperature two-point retarded correlation functions 
of the stress-energy tensor can be identified with the 
quasinormal frequencies of the gravitational perturbations in the background string theory geometry
\cite{set}. Strictly speaking, such  an identification has been made in the supergravity approximation 
to  gauge theory-string theory correspondence, but as it is derived from the standard prescription for the 
computation of the correlation functions \cite{cf1,cf2,ss}, we expect it to be valid beyond 
the supergravity approximation. In this paper we extract $\a'$-corrected transport coefficients \eqref{n4res} 
from the $\a'$-corrected dispertion relation for the lowest shear quasinormal mode \eqref{diffpole} 
and the lowest sound wave quasinormal mode  \eqref{sounddisp} in the $\a'$-corrected black 3-brane geometry
\cite{gkt,pt}.
 
We start with the  tree level type IIB low-energy effective action in ten dimensions taking into account the 
leading order string corrections \cite{cor1,cor2,cor3,cor4}
\begin{equation}
I=  \frac{1}{ 16\pi G_{10}} \int d^{10} x \sqrt {-g}
\ \bigg[ R - {\frac 12} (\partial \phi)^2 - \frac{1}{4 \cdot 5!}   (F_5)^2  +...+ 
\  \gamma \ e^{- {\frac 3 2} \phi}  W + ...\bigg]   \  ,
\eqlabel{aaa}
\end{equation}
$$ \ \ \ \ \ \   
  \gamma= { \frac 18} \zeta(3)(\alpha')^3 \ , 
$$  
where 
\begin{equation}
W =  C^{hmnk} C_{pmnq} C_{h}^{\ rsp} C^{q}_{\ rsk} 
 + {\frac 12}  C^{hkmn} C_{pqmn} C_h^{\ rsp} C^{q}_{\ rsk}\  . 
\label{rrrr}
\end{equation}
As in \cite{gkt,bls} we assume that in a chosen scheme self-dual $F_5$ form does not receive 
order $(\a')^3$
corrections.  In \eqref{aaa} ellipses stand for other fields not essential for the present analysis. 

We represent  ten dimensional background geometry describing $\gamma$-corrected black 3-branes 
by the following ansatz
\begin{equation}
\begin{split}
ds_{10}^2=&g_{\mu\nu}^{(0)}\ dx^\mu dx^\nu+c_4^2\left(dS^5\right)^2\\
\equiv&-c_1^2 dt^2+c_2^2\left(dx^2+dy^2+dz^2\right)+c_3^2dr^2+c_4^2\left(dS^5\right)^2  \,,
\end{split}
\eqlabel{g10b}
\end{equation}
where $c_i=c_i(r)$ and $\left(dS^5\right)^2$ is a metric 
on a round five-sphere of unit radius. For the dilaton we assume $\phi=\phi(r)$ and for the five-form 
\begin{equation}
F_5=\calf_5+\star\calf_5\,,\qquad \calf_5=-4\ dvol_{S^5}\,.
\eqlabel{5form}
\end{equation}
In \eqref{5form} the 5-form flux is chosen in such a way that $\gamma=0$ solution corresponds to $c_4=1$.
To leading order in $\gamma$, solution can be written explicitly \cite{gkt,pt}
\begin{equation}
\begin{split}
c_1=&r \left(1-\frac{r_0^4}{r^4}\right)^{1/2} e^{-\ft 53 \nu}\left(1+a+4b\right)\,,\\
c_2=&r e^{-\ft 53 \nu}\,,\\
c_3=&\frac{1}{r\left(1-\frac{r_0^4}{r^4}\right)^{1/2}}\  e^{-\ft 53 \nu}\left(1+b\right)\,,\\
c_4=&e^{\nu}\,,
\end{split}
\eqlabel{defm}
\end{equation} 
where to order $\calo(\gamma^2)$ 
\begin{equation}
\begin{split}
a=&-\g\ \frac{15r_0^4}{2r^4}\left(25 \frac{r_0^4}{r^4}-79 \frac{r_0^8}{r^8}+25\right)\,,\\
b=&\g\ \frac{15r_0^4}{2r^4}\left(5 \frac{r_0^4}{r^4}-19 \frac{r_0^8}{r^8}+5\right)\,,\\
\nu=&\g\ \frac{15r_0^8}{32r^8}\left(1+\frac{r_0^4}{r^4}\right)\,.
\end{split}
\end{equation}
The dilaton $\phi$ also receives $\g$ corrections, $\phi\propto \gamma$ \cite{gkt}. 
It is easy to see that to order $\calo(\g)$ gravitational perturbations do not mix with the dilaton 
perturbation; moreover to study gravitational perturbations we can consistency set $\phi=0$.
The Hawking temperature corresponding to the metric \eqref{g10b}
is \cite{gkt}
\begin{equation}
T=T_0 \left(1+15\g\right)\equiv \frac{r_0}{\pi}\left(1+15\g\right)\,.
\eqlabel{temp}
\end{equation}

Next, consider  perturbation of the five dimensional metric $g^{(0)}_{\mu\nu}$ \eqref{g10b}
\begin{equation}
g^{(0)}_{\mu\nu}\to g^{(0)}_{\mu\nu}+h_{\mu\nu}\,,
\eqlabel{hder}
\end{equation} 
where it will be sufficient to assume that 
\begin{equation}
h_{\mu\nu}=h_{\mu\nu}(t,z,r)=e^{-i\w t+i q z}\ \hh_{\mu\nu}(r)\,.
\eqlabel{hh}
\end{equation}
With the metric perturbation ansatz \eqref{hh} we have $O(2)$ rotational symmetry in 
the $xy$ plane. The latter symmetry guarantees that at the linearized level the following sets
of fluctuations decouple from each other \cite{ne2,set}
\begin{equation}
\begin{split}
\{h_{xy}\},\qquad \{h_{xx}-h_{yy}\}\,,
\end{split}
\eqlabel{set1}
\end{equation}
\begin{equation}
\begin{split}
\{h_{tx},\ h_{xz},\ h_{xr}\}\,,\qquad \{h_{ty},\ h_{yz}\,,\ h_{yr}\}\,,
\end{split}
\eqlabel{set2}
\end{equation}
\begin{equation}
\begin{split}
\{h_{tt},\ h_{tz},\ h_{tr},\ h_{aa}\equiv h_{xx}+h_{yy},\  h_{zz},\ h_{zr},\ h_{rr}\}\,.
\end{split}
\eqlabel{set3}
\end{equation}
Scalar channel fluctuations \eqref{set1} were studied in \cite{bls} leading (using the Kubo relation \eqref{kubo})
to the following prediction for the shear viscosity to the entropy density ratio
\begin{equation}
\frac{\eta}{s}=\frac{1}{4\pi} (1+135\g)\,.
\eqlabel{etasr}
\end{equation}
In this paper we study shear channel \eqref{set2}, and the sound channel \eqref{set3} fluctuations. 
Effective action for the fluctuations \eqref{set2} and \eqref{set3} can be obtained by expanding 
the supergravity action \eqref{aaa} around the background\footnote{
There is a subtlety in evaluating the action with a self-dual 5-form background. The correct way to do this is 
to assume that  $F_5$ has components only along $S^5$ and double that contribution in the 10d effective action \cite{dg}.} 
\eqref{g10b} to quadratic order in $h_{\mu\nu}$. 
Though we can always choose the gauge 
\begin{equation}
h_{tr}=h_{xr}=h_{yr}=h_{zr}=h_{rr}=0\,,
\eqlabel{gauge}
\end{equation}
doing so on the level of the effective action for the fluctuations 
would lead to missing important constraints, \ie, equations of motion coming from the 
variation of the action with respect to $\{h_{tr},\ h_{xr},\ h_{yr},\ h_{zr},\ h_{rr}\}$.
As we explicitly demonstrate on a simple example in the next section 
these constraint equations are crucial in decoupling gauge invariant fluctuations. 
Rather, the correct way is to impose the gauge fixing \eqref{gauge} 
on the level of equations of motion for the fluctuations.   

Without loss of generality, for the shear channel we consider metric perturbations 
$\{h_{tx},\ h_{xz},\ h_{xr}\}$. Imposing the gauge condition $h_{xr}$=0 on the equations 
of motion and  further introducing 
\begin{equation}
\hh_{tx}(r)=r^2\ H_{tx}(r)\,,\qquad \hh_{xz}(r)=r^2\ H_{xz}(r)\,,
\eqlabel{Hshear}
\end{equation}
we find that the shear channel gauge invariant combination \cite{set}
\begin{equation}
Z_{shear}=q H_{tx}+\w H_{xz}
\eqlabel{zshe}
\end{equation}
decouples to order $\calo(\g)$. 
The spectrum of quasinormal modes is determined \cite{set} by imposing the 
incoming wave boundary condition at the horizon $r\to r_0+$,
and the Dirichlet condition at the boundary $r\to +\infty$ in the $\g$-deformed
black 3-brane geometry \eqref{g10b} on $Z_{shear}$. The main steps of the computation are discussed in 
section 4.1. For the lowest shear quasinormal mode (in the hydrodynamic approximation)
we find to order $\calo(\g)$
\begin{equation}
\ww=-i\ \Gamma_\eta\ \qq^2+\calo(\qq^3)\,,
\eqlabel{shearres}
\end{equation} 
where 
\begin{equation}
\Gamma_\eta=\frac 12 +60\g+\calo(\g^2)\,,
\eqlabel{defgeta}
\end{equation}
and we additionally introduced\footnote{As will be clear from the discussion in section 4, $\ww$ 
and $\qq$ are the natural dimensionless parameters describing quasinormal modes.} 
\begin{equation}
\ww=\frac{\w}{2\pi T_0}\,,\qquad \qq=\frac{q}{2\pi T_0}\,.
\eqlabel{wwqqdef}
\end{equation}
From \eqref{ddef}, \eqref{temp}, \eqref{defgeta}, \eqref{wwqqdef} we find 
\begin{equation}
\frac{\eta}{s}=T \cald= T \times \frac{1}{2\pi T_0}\ \times \Gamma_\eta=\frac{1}{4\pi}\left(1+
135\g +\calo(\g^2)\right)\,,
\eqlabel{shearfinal}
\end{equation} 
in precise agreement with \eqref{etasr} reported in \cite{bls}.

There is additional subtlety in computing the lowest quasinormal frequency in the 
sound channel.  Similar to  the shear channel\footnote{The main computational steps are presented in section 4.2.}
we first derive from the effective action for the fluctuations equations of motion, and after that impose
the gauge condition 
\begin{equation}
h_{tr}=h_{zr}=h_{rr}=0\,.
\eqlabel{gauges}
\end{equation}
As explained in \cite{bbs}, for a general five-dimensional 
Einstein frame background geometry with the metric 
\begin{equation}
d\hs_5^2=-\hc_1^2\ dt^2+\hc_2^2\ (dx^2+dy^2+dz^2)+\hc_3^2\ dr^2\,,
\eqlabel{5dm}
\end{equation}
with $\hc_i=\hc_i(r)$,
in the gauge \eqref{gauges}, and reparameterizing metric perturbations \eqref{hh} as 
\begin{equation}
\hh_{tt}=\hc_1^2\ H_{tt}\,, \qquad \hh_{tz}=\hc_2^2\ H_{tz}\,,\qquad \hh_{aa}=\hc_2^2\ H_{aa}\,,
\qquad \hh_{zz}=\hc_2^2\ H_{zz}\,,   
\eqlabel{ginvc}
\end{equation}
the gauge invariant gravitational perturbation is given by 
\begin{equation}
Z_{sound}=4\ \frac q\w\ H_{tz}+2\ H_{zz}- H_{aa}\ \left(1-\frac{q^2}{\w^2}\frac{\hc_1'\hc_1}{\hc_2'\hc_2}\right)
+2\ \frac{q^2}{\w^2} \frac{\hc_1^2}{\hc_2^2}\ H_{tt}\,,
\eqlabel{zs}
\end{equation}
and thus (in the absence of matter sector) must have decoupled equation of motion.
In the absence of $\g$-corrections, the ten-dimensional Einstein frame reduces to the 
five-dimensional Einstein frame, so in defining $Z_{sound}$ we can simply take $\hc_i=c_i$. 
This is no longer the case with $\g\ne 0$, as in this case the $S^5$ warp factor $\propto c_4^2$ 
is no longer constant. Indeed, we find that defining $Z_{sound}$ as in \eqref{ginvc} produces decoupled 
equation of motion only after $\hc_i$ are rescaled as appropriate for the five-dimensional Einstein frame,
namely
\begin{equation}
\hc_i=c_4^{5/3}\ c_i\,.
\eqlabel{5deins}
\end{equation}
Again, the spectrum of quasinormal modes is determined  by imposing the 
incoming wave boundary condition at the horizon,
and the Dirichlet condition at the boundary in the $\g$-deformed
background geometry \eqref{g10b} on $Z_{sound}$. 
For the lowest shear quasinormal mode (in the hydrodynamic approximation)
we find to order $\calo(\g)$
\begin{equation}
\ww=c_s\ \qq-i\ \Gamma_{sound}\ \qq^2+\calo(\qq^3)\,,
\eqlabel{soundres}
\end{equation} 
where 
\begin{equation}
\begin{split}
c_s=&\frac{1}{\sqrt{3}}+\calo(\g^2)\,,
\cr
\Gamma_\eta=&\frac 13 +40\g+\calo(\g^2)\,.
\end{split}
\eqlabel{csgs}
\end{equation}
Given \eqref{sounddisp}, \eqref{temp}, \eqref{wwqqdef}, \eqref{shearfinal} we conclude from 
\eqref{csgs} that the bulk viscosity of the strongly coupled $\caln=4$ plasma is 
\begin{equation}
\zeta=\calo(\g^2)\,.
\eqlabel{zetadef}
\end{equation}
Of course, as finite $\g$-corrections translate into 't Hooft coupling corrections on the gauge theory 
side of the Maldacena correspondence, from the field theory perspective (given that the latter is conformal) 
we immediately conclude that $c_s^2=\ft 13$ and $\zeta=0$ independent of the 't Hooft coupling. 
We showed here that the dual string theory description reproduces this fact as well, albeit in a highly nontrivial 
fashion which is  moreover consistent with shear viscosity computation \eqref{shearfinal} and alternative 
analysis in \cite{bls}.

\section{Diffusion constant of the black 3-branes hydrodynamics: the effective action 
approach}
Consider the shear channel gravitational perturbations $\{h_{tx}, h_{xz}, h_{xr}\}$ 
in the absence of $\a'$ corrections, \ie, setting $\g=0$. 
This was previously discussed in \cite{ne2}, where equations of motion for the fluctuations 
 $\{h_{tx}, h_{xz}\}$ in the gauge $h_{xr}=0$ were derived as perturbation of the full type IIB supergravity 
equations of motion. Such an approach becomes technically very difficult in the presence of $\g$ corrections:
one needs to derive equations of motion for the deformed effective type IIB supergravity action \eqref{aaa}. 
In the latter case we find it much easier to derive first the effective action describing the 
fluctuations, and then derive the equations of motion from this action. The effective action 
for the fluctuations can be obtained by simply evaluating \eqref{aaa} to quadratic order in metric perturbations
\eqref{hder}. The important point we want to stress here is that the gauge fixing condition 
$h_{xr}=0$ can not be imposed on the level of action. If we do this, we obtain two second order ODE's (coming from 
variation of the action with respect to  $\{h_{tx}, h_{xz}\}$  )
\begin{equation}
\begin{split}
0=&H_{tx}''-\frac 1x H_{tx}'-\frac{\qq}{(1-x^2)^{3/2}}\ \biggl(\ww H_{xz}+\qq H_{tx}\biggr)\,,
\cr
0=&H_{xz}''+\frac 1x H_{xz}'+\frac{\ww}{x^2(1-x^2)^{3/2}}\ \biggl(\ww H_{xz}+\qq H_{tx}\biggr)\,,
\end{split}
\eqlabel{2diff}
\end{equation}
where $H_{\cdots}=H_{\cdots}(x)$, and all the derivatives are with respect to 
\begin{equation}
x\equiv \left(1-\frac{r_0^4}{r^4}\right)^{1/2}\,.
\eqlabel{xdef}
\end{equation}
It is easy to see that given \eqref{xdef} equation of motion for 
\begin{equation}
Z_{shear}(x)=\qq H_{tx}(x)+\ww H_{xz}(x)
\end{equation}
does not decouple. On the other hand, if we impose the gauge fixing condition $h_{xr}=0$
on the level of equations of motion, we obtain an extra constraint equation coming  from the variation of the 
effective action for the fluctuations with respect to $h_{xr}$ 
\begin{equation}
0=\ww H_{tx}'+\qq x^2 H_{xz}'\,.
\eqlabel{3diff}
\end{equation}
Notice that \eqref{3diff} is consistent with \eqref{2diff}. Given \eqref{3diff} we can now obtain the 
decoupled equation of motion for $Z_{shear}$
\begin{equation}
0=Z_{shear}''+\frac{x^2\qq^2+\ww^2}{x(\ww^2-x^2 \qq^2)}\ Z_{shear}'+\frac{\ww^2-x^2 \qq^2}{x^2(1-x^2)^{3/2}}\ 
Z_{shear}\,.
\eqlabel{zseq}
\end{equation}
The incoming boundary condition at the horizon ($x\to 0_+$) implies that 
\begin{equation}
Z_{shear}(x)=x^{-i\ww} z_{shear}(x)\,,
\eqlabel{diffbh}
\end{equation}
where $z_{shear}(x)$ is regular at the horizon. Without loss of generality we can assume 
\begin{equation}
z_{shear}\bigg|_{x\to 0_+}=1\,,
\eqlabel{diffbh1}
\end{equation}
the spectrum of quasinormal frequencies is then determined by imposing a Dirichlet condition at the boundary 
\cite{set}
\begin{equation}
z_{shear}\bigg|_{x\to 1_-}=0\,.
\eqlabel{diffbh12}
\end{equation}
In the hydrodynamic approximation ($\ww\ll 1$ and $\qq\ll 1$) the solution can be written in the ansatz 
\begin{equation}
z_{shear}=z_{shear}^{(0)}+i\ \qq z_{shear}^{(1)}+\calo(\qq^2)\,,
\eqlabel{zd1}
\end{equation}
where $z_{shear}^{(0)},\ z_{shear}^{(1)}$ are invariant under the scaling $\ww\to \lambda \ww,\ \qq\to \lambda \qq$
with constant $\lambda$. Substituting \eqref{zd1} into \eqref{zseq}, and  we find \cite{ne2,set}
\begin{equation}
z_{shear}^{(0)}=1\,,\qquad z_{shear}^{(1)}=\frac 12 \frac{\qq}{\ww}\ x^2\,,
\eqlabel{zd2} 
\end{equation}
which from \eqref{diffbh12} determines the lowest shear quasinormal frequency as \cite{set}
\begin{equation}
\ww=-i\ \frac 12\ \qq^2 \,.
\end{equation}

\section{Transport properties of black 3-branes at $\calo((\a')^3)$ order}
The general computational scheme for deriving equations of motion for the metric perturbation 
and decoupling the gauge invariant combinations for these perturbations is explained in section 2.  
The analysis is straightforward, though quite tedious. As in the previous section,  computations 
are simplified using the radial coordinate $x$, defined by \eqref{xdef}. Both for the 
shear mode \eqref{zshe} and the 
sound  mode \eqref{zs} gauge invariant combination of metric perturbations we find that the corresponding 
equations  of motion decouple. These equations can be expanded perturbatively 
in $\g$, provided we introduce
\begin{equation}
\begin{split}
Z_{shear}=&Z_{shear,0}+\g\ Z_{shear,1}+\calo(\g^2)\,,\\
Z_{sound}=&Z_{sound,0}+\g\ Z_{sound,1}+\calo(\g^2)\,.
\end{split}
\end{equation} 
The incoming wave boundary conditions are set up at the level of the leading  order in $\g$, thus it is not a 
surprise that a natural dimensionless frequency $\ww$ and a momentum $\qq$ are introduced (see eq.~\eqref{wwqqdef}) 
with respect to $T_0$, rather than the $\a'$-corrected Hawking temperature $T$ of the black branes \eqref{temp}.  

\subsection{Shear quasinormal mode}
For the shear channel fluctuations we find 
\begin{equation}
\begin{split}
0=&Z_{shear,0}''+\frac{x^2\qq^2+\ww^2}{x(\ww^2-x^2 \qq^2)}\ Z_{shear,0}'+\frac{\ww^2-x^2 \qq^2}{x^2(1-x^2)^{3/2}}\ 
Z_{shear,0}\,,\\
0=&Z_{shear,1}''+\frac{x^2\qq^2+\ww^2}{x(\ww^2-x^2 \qq^2)}\ Z_{shear,1}'+\frac{\ww^2-x^2 \qq^2}{x^2(1-x^2)^{3/2}}\ 
Z_{shear,1}+J_{shear,0}\,,
\end{split}
\eqlabel{zsea1}
\end{equation}
where the source  $J_{shear,0}$ is a functional of the zero's order shear mode $Z_{shear,0}$
\begin{equation}
\begin{split}
J_{shear,0}=&\calc_{shear}^{(4)}  \frac{d^4 Z_{shear,0}}{d x^4}+\calc_{shear}^{(3)}
\ \frac{d^3 Z_{shear,0}}{d x^3}+\calc_{shear}^{(2)}
\  \frac{d^2 Z_{shear,0}}{d x^2}+\calc_{shear}^{(1)}\ \frac{d Z_{shear,0}}{d x}\\
&+\calc_{shear}^{(0)}\ Z_{shear,0}\,.
\end{split}
\eqlabel{sourceshear}
\end{equation}
The coefficients $\calc_{shear}^{(i)}$ are given explicitly in appendix A.
In the hydrodynamic approximation we look for the solution for $Z_{shear,i}$ in the following ansatz
\begin{equation}
\begin{split}
Z_{shear,0}=&x^{-i\ww}\ \left(z_{shear,0}^{(0)}+i\qq z_{shear,0}^{(1)}+\calo(\qq^2)\right)\,,\\
Z_{shear,1}=&x^{-i\ww}\ \left(z_{shear,1}^{(0)}+i\qq z_{shear,1}^{(1)}+\calo(\qq^2)\right)\,,
\end{split}
\eqlabel{sss1}
\end{equation} 
where $z_{shear,i}^{(j)}$ are regular at the horizon, and satisfy the following boundary conditions
\begin{equation}
z_{shear,0}^{(0)}\bigg|_{x\to 0_+}=1\,,\qquad z_{shear,0}^{(1)}\bigg|_{x\to 0_+}=z_{shear,1}^{(0)}\bigg|_{x\to 0_+}=
z_{shear,1}^{(1)}\bigg|_{x\to 0_+}=0\,.
\eqlabel{sss2}
\end{equation}
Explicit solution of \eqref{zsea1} subject to boundary conditions \eqref{sss2} takes form
\begin{equation}
\begin{split}
z_{shear,0}^{(0)}=1\,,\qquad z_{shear,0}^{(1)}=\frac 12 \frac {\qq}{\ww} x^2\,,
\end{split}
\eqlabel{sss3}
\end{equation}
\begin{equation}
\begin{split}
z_{shear,1}^{(0)}=&\frac{25}{16}x^2\left(x^4-4x^2+5\right)\,,
\\
z_{shear,1}^{(1)}=&-\frac{1}{32\qq \ww}x^2\biggl(\qq^2\left(-240-1565 x^2-860 x^4+695x^6\right)\\
&+16\ww^2
\left(594-264x^2+43x^4\right)\biggr)\,.
\end{split}
\eqlabel{sss4}
\end{equation}
Imposing the Dirichlet condition on $x^{i\ww} Z_{shear,0}$ at the boundary determines the 
lowest shear quasinormal frequency
\eqref{shearres}.

\subsection{Sound wave quasinormal mode}
For the sound channel fluctuations we find 
\begin{equation}
\begin{split}
0=&Z_{sound,0}''-\frac{(3x^2-2)\qq^2+3\ww^2}{x(-3\ww^2+(x^2+2) \qq^2)}\ Z_{sound,0}'
\\
&-\frac{\qq^4 x^2(x^2+2)-2\qq^2\ww^2(2x^2+2)-4x^2(1-x^2)^{3/2}\qq^2+3\ww^4}{x^2(1-x^2)^{3/2}((x^2+2)\qq^2-3\ww^2)}\ 
Z_{sound,0}\,,\\
0=&Z_{sound,1}''-\frac{(3x^2-2)\qq^2+3\ww^2}{x(-3\ww^2+(x^2+2) \qq^2)}\ Z_{sound,1}'
\\
&-\frac{\qq^4 x^2(x^2+2)-2\qq^2\ww^2(2x^2+2)-4x^2(1-x^2)^{3/2}\qq^2+3\ww^4}{x^2(1-x^2)^{3/2}((x^2+2)\qq^2-3\ww^2)}\ 
Z_{sound,1}+J_{sound,0}\,,
\end{split}
\eqlabel{szsea1}
\end{equation}
where the source  $J_{sound,0}$ is a functional of the zero's order sound mode $Z_{sound,0}$
\begin{equation}
\begin{split}
J_{sound,0}=&\calc_{sound}^{(4)}  \frac{d^4 Z_{sound,0}}{d x^4}+\calc_{sound}^{(3)}
\ \frac{d^3 Z_{sound,0}}{d x^3}+\calc_{sound}^{(2)}
\  \frac{d^2 Z_{sound,0}}{d x^2}+\calc_{sound}^{(1)}\ \frac{d Z_{sound,0}}{d x}\\
&+\calc_{sound}^{(0)}\ Z_{sound,0}\,.
\end{split}
\eqlabel{sourcesound}
\end{equation}
The coefficients $\calc_{sound}^{(i)}$ are given explicitly in appendix B.
In the hydrodynamic approximation we look for the solution for $Z_{sound,i}$ in the following ansatz
\begin{equation}
\begin{split}
Z_{sound,0}=&x^{-i\ww}\ \left(z_{sound,0}^{(0)}+i\qq z_{sound,0}^{(1)}+\calo(\qq^2)\right)\,,\\
Z_{sound,1}=&x^{-i\ww}\ \left(z_{sound,1}^{(0)}+i\qq z_{sound,1}^{(1)}+\calo(\qq^2)\right)\,,
\end{split}
\eqlabel{ssss1}
\end{equation} 
where $z_{sound,i}^{(j)}$ are regular at the horizon, and satisfy the following boundary conditions
\begin{equation}
z_{sound,0}^{(0)}\bigg|_{x\to 0_+}=1\,,\qquad z_{sound,0}^{(1)}\bigg|_{x\to 0_+}=z_{sound,1}^{(0)}\bigg|_{x\to 0_+}=
z_{sound,1}^{(1)}\bigg|_{x\to 0_+}=0\,.
\eqlabel{ssss2}
\end{equation}
Explicit solution of \eqref{szsea1} subject to boundary conditions \eqref{ssss2} takes form
\begin{equation}
\begin{split}
z_{sound,0}^{(0)}=\frac{3\ww^2+(x^2-2)\qq^2}{3\ww^2-2\qq^2},\qquad z_{sound,0}^{(1)}=\frac {2\ww\qq x^2}
{3\ww^2-2\qq^2}\,,
\end{split}
\eqlabel{ssss3}
\end{equation}
\begin{equation}
\begin{split}
z_{sound,1}^{(0)}=&\frac{5x^2}{16(3\ww^2-2\qq^2)^2}\biggl(\qq^4\left(2404+446x^2-4164x^4+2006x^6\right)
\\
&-3\ww^2\qq^2\left(1588+183x^2-2072x^4+1003x^6\right)+45\ww^4\left(5-4x^2+x^4\right)
\biggr)\,,
\\
z_{sound,1}^{(1)}=&\frac{\ww x^2}{8\qq (3\ww^2-2\qq^2)^2}\biggl(\qq^4\left(-13344+5846x^2-4520x^4+1734x^6\right)
\\
&-3\ww^2\qq^2\left(-9744+5035x^2-2604x^4+867x^6\right)\\
&-36\ww^4\left(594-264x^2+43x^4\right)
\biggr)\,.
\end{split}
\eqlabel{ssss4}
\end{equation}
Imposing the Dirichlet condition on $x^{i\ww} Z_{sound,0} $ at the boundary 
determines the lowest sound quasinormal frequency
\eqref{soundres}.

\section*{Appendix}
\appendix

\section{Coefficients of $J_{shear,0}$}
\begin{equation}
\begin{split}
\calc_{shear}^{(4)}=&45(1-x^2)^4
\end{split}
\eqlabel{csh4}
\end{equation}
\begin{equation}
\begin{split}
\calc_{shear}^{(3)}=&90(1-x^2)^3\frac{({\qq}^2(7x^2+1)-8{\ww}^2)x}{-{\qq}^2x^2+{\ww}^2}
\end{split}
\eqlabel{csh3}
\end{equation}
\begin{equation}
\begin{split}
&\calc_{shear}^{(2)}=-\frac{1}{8x^2(-{\qq}^2x^2+{\ww}^2)}
\biggl(5(1-x^2)^2({\qq}^2x^2(2803x^4+1018x^2+216)-{\ww}^2(72
\\
&+3811x^4+154x^2))+16(1-x^2)^{5/2}({\qq}^2x^2-{\ww}^2)(45{\ww}^2+83{\qq}^2x^2)\biggr)
\end{split}
\eqlabel{csh2}
\end{equation}
\begin{equation}
\begin{split}
&\calc_{shear}^{(1)}=-\frac{1}{8x^3(-{\qq}^2x^2+{\ww}^2)^2}\biggl(-5{\qq}^4x^4(1-x^2)(1105x^6-2411x^4-802x^2-216)\\
&-4{\ww}^2{\qq}^2x^2(-8792x^6+1467x^8+8045x^4+60x^2-1080)
-{\ww}^4(-13479x^4
+3353x^8\\
&+2930x^2+5876x^6+2520)+16(1-x^2)^{3/2}(83x^6(4x^2+1){\qq}^6
-{\ww}^2{\qq}^4x^4(497x^2-372)
\\
&-{\qq}^2{\ww}^4x^2(162x^2+353)+45{\ww}^6(3x^2+2))\biggr)
\end{split}
\eqlabel{csh1}
\end{equation}
\begin{equation}
\begin{split}
&\calc_{shear}^{(0)}=-\frac{1}{8x^4(-{\qq}^2x^2+{\ww}^2)}\biggl(152{\qq}^6x^6(-1+x^2)-1480{\ww}^2{\qq}^4x^4(-1+x^2)\\
&+4x^2{\qq}^2(-242{\ww}^4+242{\ww}^4x^2-75x^8+100x^6)+10{\ww}^2(-36{\ww}^4+36{\ww}^4x^2
+45x^8\\
&-80x^6+25x^4)+(-5{\qq}^4x^6(-831x^2+274+677x^4)
+2{\ww}^2{\qq}^2x^2(-1858x^2+4453x^6\\
&-5295x^4+3600)-(4320+1250x^2+6097x^6
-10467x^4){\ww}^4)(1-x^2)^{-1/2}\biggr)
\end{split}
\eqlabel{csh0}
\end{equation}

\section{Coefficients of $J_{sound,0}$}
\begin{equation}
\begin{split}
\calc_{sound}^{(4)}=&\frac{2 {\qq}^2 (183 x^2+29)+333 {\ww}^2}{3(2 {\qq}^2 (x^2-1)+3 {\ww}^2)} (1-x^2)^4 
\end{split}
\eqlabel{cso4}
\end{equation}
\begin{equation}
\begin{split}
&\calc_{sound}^{(3)}= \frac{2(x^2-1)^3}{3x ({\qq}^2 (x^2+2)-3 {\ww}^2) (2 {\qq}^2 (x^2-1)+3 {\ww}^2)^2}  \biggl(
4 {\qq}^6 (x^2-1) (872 x^6+2486 x^4\\
&-567 x^2-58)-6 {\ww}^2 {\qq}^4 (557 x^6-7955 x^4+2386 x^2-106)-9 {\ww}^4 {\qq}^2 
(4237 x^4-3875 x^2\\
&+386)-2997 {\ww}^6 (9 x^2-1)\biggr)
\end{split}
\eqlabel{cso3}
\end{equation}
\begin{equation}
\begin{split}
&\calc_{sound}^{(2)}=\frac{1}{(24 x^2 ({\qq}^2 (x^2+2)-3 {\ww}^2))(2 {\qq}^2 (x^2-1)+3 {\ww}^2)^3}
 \biggl( 8 (17357 x^{10}+64126 x^8\\
&-125343 x^6+33528 x^4+4468 x^2+464) (x^2-1)^3 {\qq}^8-12 {\ww}^2 (12419 x^{10}
-416676 x^8\\
&+639279 x^6-205814 x^4-18792 x^2+384) (x^2-1)^2 {\qq}^6
-18 {\ww}^4 (x^2-1) (183535 x^{10}\\
&-1069144 x^8+1301579 x^6-428618 x^4+14112 x^2
+3936) {\qq}^4-27 {\ww}^6 (x^2-1) (359001 x^8\\
&-732179 x^6+440634 x^4-71592 x^2
-4864) {\qq}^2-243 {\ww}^8 (x^2-1) (22327 x^6-28997 x^4\\
&+6974 x^2+296)- 16 (2 {\qq}^2 (x^2-1)+3 {\ww}^2) ({\qq}^2 (x^2+2)-3 {\ww}^2) (1-x^2)^{5/2} (4 x^2 (x^2-1)\\
&\times (125 x^2-87) {\qq}^6
+4 {\ww}^2 (273 x^4-86 x^2+29) {\qq}^4-3 {\ww}^4 (159 x^2-164) {\qq}^2-999 {\ww}^6)\biggr)
\end{split}
\eqlabel{cso2}
\end{equation}
\begin{equation}
\begin{split}
&\calc_{sound}^{(1)}=\frac{1}{24 x^3 (2 {\qq}^2 (x^2-1)+3 {\ww}^2)^3 ({\qq}^2 (x^2+2)-3 {\ww}^2)^3} 
\biggl( 8 (461207 x^{14}+1124342 x^{12}\\
&-1565613 x^{10}-408216 x^8+1082544 x^6-675120 x^4+31312 x^2-1856) (x^2-1)^3 {\qq}^{12}\\
&-12 {\ww}^2 (234845 x^{14}-6242508 x^{12}+5011337 x^{10}+4244306 x^8-6922564 x^6\\
&+3818200 x^4-275392 x^2+2176) (x^2-1)^2 {\qq}^{10}-18 {\ww}^4 (x^2-1) (3115141 x^{14}\\
&-10599760 x^{12}+3749965 x^{10}+13178414 x^8-17763088 x^6+9000248 x^4-777360 x^2\\
&-16960) {\qq}^8-27 {\ww}^6 (-52480+19140582 x^8+10819632 x^4-7115740 x^{12}+3171763 x^{14}\\
&-23084512 x^6-1929885 x^{10}-981760 x^2) {\qq}^6-81 {\ww}^8 (313560 x^2+591463 x^{12}\\
&-1966197 x^8-476182 x^{10}-2942744 x^4+4503220 x^6+30880) {\qq}^4-243 {\ww}^{10} (-8416\\
&+96733 x^{10}-43280 x^2+303614 x^4-158080 x^8-215771 x^6) {\qq}^2-2187 {\ww}^{12} (7105 x^8\\
&-7980 x^6+296+426 x^2+1353 x^4)
-16 ({\qq}^2 (x^2+2)-3 {\ww}^2) (2 {\qq}^2 (x^2-1)+3 {\ww}^2)\\
&\times  (1-x^2)^{3/2} ( 4 x^2 (x^2-1) (2863 x^8+5423 x^6-8425 x^4-1988 x^2-348) {\qq}^{10}\\
&-2 {\ww}^2 (4979 x^{10}-83715 x^8+87128 x^6+16642 x^4+3624 x^2+232) {\qq}^8-3 {\ww}^4 (192\\
&+38586 x^8-110297 x^6-4976 x^2-22460 x^4) {\qq}^6-9 {\ww}^6 (20610 x^6+4957 x^4-148 x^2\\
&-984) {\qq}^4-27 {\ww}^8 (206 x^4+1561 x^2+608) {\qq}^2+8991 {\ww}^{10} (1+4 x^2))\biggr)
\end{split}
\eqlabel{cso1}
\end{equation}
\begin{equation}
\begin{split}
&\calc_{sound}^{(0)}=\frac{1}{24 x^4 (2 {\qq}^2 (x^2-1)+3 {\ww}^2)^3 ({\qq}^2 (x^2+2)-3 {\ww}^2)^3} \biggl(
 -64 x^4 (643 x^2-779)
 (x^2-1)^3\\
& \times(x^2+2)^3 {\qq}^{16}+32 x^2 {\ww}^2 (5747 x^6-31591 x^4+25688 x^2+696) (x^2-1)^2 (x^2+2)^2 {\qq}^{14}\\
&-16 (x^2-1) (343159 x^{18}+359756 x^{16}-2430770 x^{14}-19905 {\ww}^4 x^{12}+1208858 x^{12}\\
&-216211 {\ww}^4 x^{10}+2130051 x^{10}
-1974730 x^8+212468 {\ww}^4 x^8+789188 {\ww}^4 x^6+317076 x^6\\
&-704624 {\ww}^4 x^4+46600 x^4-43376 {\ww}^4 x^2+928 {\ww}^4) {\qq}^{12}+24 {\ww}^2 (x^2-1) (385565 x^{16}\\
&-4651541 x^{14}+3705393 x^{12}+6897979 x^{10}-73557 {\ww}^4 x^{10}-7874224 x^8+121796 {\ww}^4 x^8\\
&+1319004 x^6+803308 {\ww}^4 x^6-855456 {\ww}^4 x^4+217824 x^4-96080 {\ww}^4 x^2+1088 {\ww}^4) {\qq}^{10}\\
&+36 {\ww}^4 (x^2-1) (2216157 x^{14}-3463628 x^{12}-8055387 x^{10}-26682 {\ww}^4 x^8+12127416 x^8\\
&-2103232 x^6-408776 {\ww}^4 x^6-547176 x^4+579864 {\ww}^4 x^4+117600 {\ww}^4 x^2+8480 {\ww}^4) {\qq}^8\\
&+54 {\ww}^6 (26240 {\ww}^4+10762468 x^8-803584 x^4+2980096 x^{12}+956891 x^{14}-462808 x^6\\
&-13364213 x^{10}+122304 {\ww}^4 x^4-290744 {\ww}^4 x^6+58560 {\ww}^4 x^2+83640 {\ww}^4 x^8) {\qq}^6\\
&-162 {\ww}^8 (15440 {\ww}^4+355529 x^{12}+1905537 x^8-2260254 x^{10}-303468 x^4+349906 x^6\\
&-2056 {\ww}^4 x^4-15120 {\ww}^4 x^6+1736 {\ww}^4 x^2) {\qq}^4-486 {\ww}^{10} (-4208 {\ww}^4+150699 x^{10}+45576 x^4\\
&-93108 x^8-111717 x^6+2684 {\ww}^4 x^2+1524 {\ww}^4 x^4) {\qq}^2+4374 {\ww}^{12} (-148 {\ww}^4-400 x^6\\
&+148 {\ww}^4 x^2+125 x^4+225 x^8)-({\qq}^2 (x^2+2)-3 {\ww}^2) (1-x^2)^{-1/2} ( 8 x^4 (117035 x^{10}\\
&+280248 x^8-600005 x^6-185858 x^4+519204 x^2-114424) (x^2-1)^2 {\qq}^{12}-4 {\ww}^2 (x^2-1) \\
&\times (403063 x^{14}-5091874 x^{12}+4346503 x^{10}+5410988 x^8-6467736 x^6+1273520 x^4\\
&+35760 x^2-7424) {\qq}^{10}-6 {\ww}^4 (10631928 x^8+2466759 x^{14}+1592283 x^{10}-7034846 x^{12}\\
&-8874092 x^6+974760 x^4+195888 x^2-1280) {\qq}^8-27 {\ww}^6 (-797920 x^8+485056 x^4\\
&+475731 x^{12}+927244 x^6-918927 x^{10}-23040-115744 x^2) {\qq}^6-27 {\ww}^8 (-269131 x^8\\
&+473495 x^{10}-1348292 x^4+908224 x^6+70400+149104 x^2) {\qq}^4-81 {\ww}^{10} (314448 x^4\\
&+56787 x^8-321287 x^6-26560-30588 x^2) \qq^2-729 {\ww}^{12} (1184+2665 x^6-4027 x^4\\
&+778 x^2))\biggr)
\end{split}
\eqlabel{cso0}
\end{equation}

\section*{Acknowledgments}
Research at Perimeter Institute is supported in part by funds from NSERC of
Canada. AB gratefully   acknowledges  support by  NSERC Discovery
grant.

\end{document}